\documentclass[pdflatex,sn-mathphys-num]{sn-jnl}

\usepackage{graphicx}%
\usepackage{multirow}%
\usepackage{amsmath,amssymb,amsfonts}%
\usepackage{amsthm}%
\usepackage{mathrsfs}%
\usepackage[title]{appendix}%
\usepackage{xcolor}%
\usepackage{textcomp}%
\usepackage{manyfoot}%
\usepackage{booktabs}%
\usepackage{algorithm}%
\usepackage{algorithmicx}%
\usepackage{algpseudocode}%
\usepackage{listings}%
\usepackage{lineno}

\theoremstyle{thmstyleone}%
\theoremstyle{thmstyletwo}%

\theoremstyle{thmstylethree}%

\begin{document}

\title[Article Title]{scMamba: A Scalable Foundation Model for Single-Cell Multi-Omics Integration Beyond Highly Variable Feature Selection
}


\author[1]{\fnm{Zhen} \sur{Yuan}}

\author[1]{\fnm{Shaoqing} \sur{Jiao}}

\author[1]{\fnm{Yihang} \sur{Xiao}}
\author*[1,2]{\fnm{Jiajie} \sur{Peng}}\email{jiajiepeng@nwpu.edu.cn}

\affil[1]{\orgdiv{AI for Science Interdisciplinary Research Center}, \orgname{School of Computer Science, Northwestern Polytechnical University}, \orgaddress{\street{No.1 Dongxiang Road}, \city{Xi'an}, \country{China}}}

\affil[2]{\orgdiv{Key Laboratory of Big Data Storage and Management, Northwestern Polytechnical University, Ministry of Industry and Information Technology}, \orgaddress{\street{No.1 Dongxiang Road}, \city{Xi'an}, \country{China}}}


\abstract{
The advent of single-cell multi-omics technologies has enabled the simultaneous profiling of diverse omics layers within individual cells. Integrating such multimodal data provides unprecedented insights into cellular identity, regulatory processes, and disease mechanisms. However, it remains challenging, as current methods often rely on selecting highly variable genes or peaks during preprocessing, which may inadvertently discard crucial biological information. Here, we present scMamba, a foundation model designed to integrate single-cell multi-omics data without the need for prior feature selection while preserving genomic positional information. scMamba introduces a patch-based cell tokenization strategy that treats genomics regions as words (tokens) and cells as sentences. Building upon the concept of state space duality, scMamba distills rich biological insights from high-dimensional, sparse single-cell multi-omics data. Additionally, our novel contrastive learning approach, enhanced with cosine similarity regularization, enables superior alignment across omics layers compared to traditional methods. Systematic benchmarking across multiple datasets demonstrates that scMamba significantly outperforms state-of-the-art methods in preserving biological variation, aligning omics layers, and enhancing key downstream tasks such as clustering, cell type annotation, and trajectory inference. Our findings position scMamba as a powerful tool for large-scale single-cell multi-omics integration, capable of handling large-scale atlases and advancing biological discovery.
}




\maketitle
\section{Introduction}\label{sec1}
The development of single-cell sequencing technologies has revolutionized our ability to detect multiple types of genomic information within individual cells, such as RNA expression, chromatin accessibility, and cell surface protein abundance. In recent years, several single-cell multi-omics technologies have been developed, enabling the simultaneous profiling of multiple omics layers at the single-cell level. For example, CITE-seq \cite{stoeckius2017simultaneous} and REAP-seq \cite{peterson2017multiplexed} allow the simultaneous measurement of RNA expression and surface protein abundance in individual cells. SHARE-seq \cite{ma2020chromatin}, SNARE-seq \cite{chen2019high}, and 10x Genomics Multiome simultaneously capture RNA expression and chromatin accessibility from the same cell. 
These technologies enable detailed characterization of cellular heterogeneity across multiple omics layers, enhancing our understanding of cellular diversity \cite{baysoy2023technological, ma2025scmfg, liu2024spatial, liu2024spatiotemporal}. Additionally, they provide valuable insights into the interplay of omics features, offering a comprehensive view of cellular regulation and interactions at single-cell resolution \cite{lim2024advances, he2024mosaic, guan2025single}.
As single-cell multi-omics datasets expand in size and complexity, there is an increasing demand for novel computational methods to effectively integrate these high-dimensional data and support reliable downstream analyses \cite{stuart2019integrative, baysoy2023technological, vandereyken2023methods}.
Integrating multi-omics data provides unprecedented insights into cellular processes, cell identification, and disease mechanisms \cite{baysoy2023technological, lim2024advances} while also facilitating the construction of large-scale single-cell multimodal atlases \cite{he2024mosaic}, enabling the full potential of publicly available multi-omics data to be realized.
%

Recently, various methods have been proposed to integrate single-cell multi-omics data. Most of them involve converting multi-omics data into a common feature space using prior knowledge and then applying batch integration methods \cite{korsunsky2019fast, hie2019efficient, lopez2018deep, lotfollahi2022mapping, xu2021probabilistic, xiong2022online}. For instance, Harmony \cite{korsunsky2019fast} employs iterative correction of principal components to effectively reduce batch effects while preserving biological variation in single-cell datasets. scVI \cite{lopez2018deep}, scArches \cite{lotfollahi2022mapping}, and SCALEX \cite{xiong2022online} leverage a variational autoencoder (VAE) \cite{kingma2013auto} framework to model the underlying distribution of single-cell data for integration. However, this explicit ``feature conversion" has been reported to cause information loss \cite{chen2019assessment} and ignore non-overlapping features \cite{ghazanfar2024stabilized}. Moreover, applying batch integration methods to single-cell multi-omics data is often inefficient, as the data distribution and sparsity levels are vastly different across omics \cite{lin2022scjoint}.

To overcome this challenge, considerable efforts have focused on integrating single-cell multi-omics data \cite{hao2021integrated, gayoso2021joint, li2022deep, welch2019single, cao2022multi, liu2022cvqvae, xiong2023scclip}.  
totalVI \cite{gayoso2021joint} is a VAE framework that jointly models paired single-cell transcriptomic and proteomic data, enabling denoising, batch correction, and the integration of multi-omic measurements within a unified latent space. scMVP \cite{li2022deep} applies a non-symmetric multi-view VAE with self-attention to integrate scRNA-seq and scATAC-seq data into a shared low-dimensional latent space for cell embedding. 
Nonetheless, these approaches generate unified representations rather than modality-specific joint embeddings, which may result in an imbalanced contribution from different modalities, often favoring scRNA-seq data \cite{chen2024integration, xiong2023scclip}.  
Alternatively, GLUE \cite{cao2022multi} is a multimodal framework that integrates single-cell RNA and ATAC sequencing data using graph variational autoencoders combined with an adversarial training strategy, leveraging prior knowledge of gene regulatory networks to enhance data integration and uncover gene regulatory mechanisms at single-cell resolution. CVQVAE \cite{liu2022cvqvae} leverages cross-trained vector-quantized variational autoencoders to efficiently integrate single-cell multi-omics data while preserving biological information and removing batch effects. scCLIP \cite{xiong2023scclip} leverages a generalized multi-modal transformer model with contrastive learning to integrate paired scRNA-seq and scATAC-seq data. 
However, to alleviate computational inefficiency, most single-cell integration models preprocess data by selecting approximately 3,000 to 5,000 highly variable genes or peaks, based on the assumption that these features capture the most informative biological signals for downstream analysis. 
While this dimensionality reduction strategy simplifies feature extraction and enhances computational efficiency, it may also exclude important yet less variable features, thereby limiting the model's ability to fully leverage the rich and complex information embedded in single-cell data.

To efficiently leverage raw features and diverse modalities in single-cell multi-omics data, we present scMamba, a foundation model inspired by the Mamba architecture \cite{dao2024transformers}, specifically designed for single-cell multi-omics integration.
Unlike traditional approaches, scMamba operates directly on single-cell data without prior selection of highly variable features, thereby capturing more comprehensive and biologically informative signals.
In addition, existing single-cell foundation models \cite{cui2024scgpt, yang2022scbert, hao2024large} treat genes as words (tokens), which ignore the genomic positional information. To fill this gap, we introduce a patch-based cell tokenization strategy that treats genomics regions as words and cells as sentences. Each genomic region represents the set of genes or chromatin accessibility peaks located within it, ordered according to their genomic coordinates. 
This strategy abstracts high-dimensional single-cell inputs into genomic regions, enabling scMamba to process tens of thousands of features efficiently. 
To efficiently handle the extremely sparse single-cell data, we design the scMamba encoder based on the state space duality algorithm \cite{dao2024transformers}, which facilitates the learning of biological insights from raw genomic features.
As most single-cell multi-omics data are unlabeled, we propose a self-supervised training strategy based on a novel contrastive learning approach, incorporating an additional cosine similarity regularization module. This approach enhances the multimodal representation learning of biological signals and further aligns cell embeddings across omics layers.
Through systematic benchmarking across multiple publicly available datasets, scMamba consistently outperforms several state-of-the-art integration methods. In particular, it achieves an average improvement of more than 10\% in the overall integration score and demonstrates robust scalability to atlas-level single-cell multi-omics data.
Moreover, scMamba generates informative cell embeddings that consistently enhance the performance of downstream tasks, including clustering, cell type annotation, and trajectory inference.

\begin{figure}[htbp]
    \centering
    \includegraphics[width=1.0\textwidth]{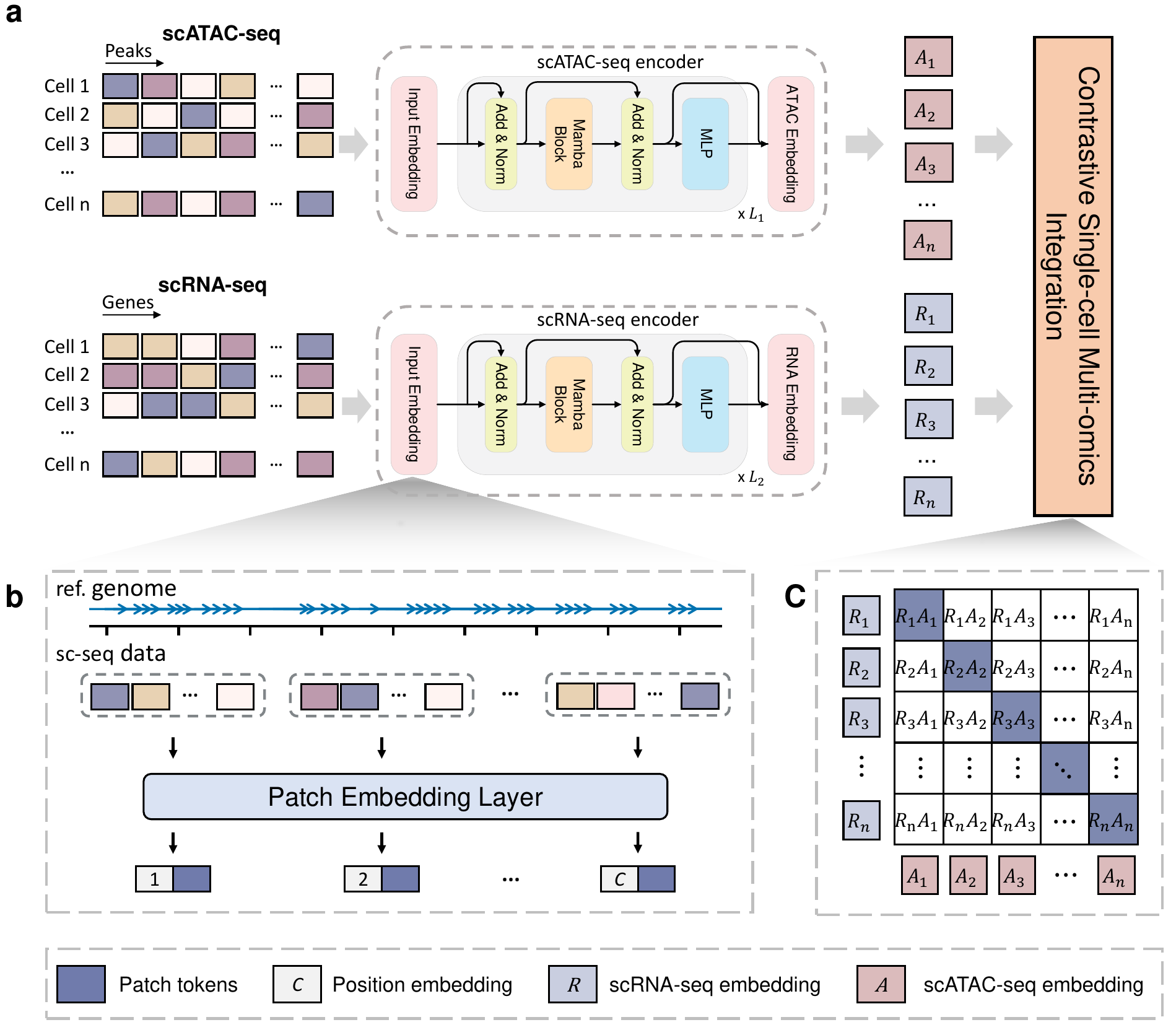}
    \caption{\textbf{Schematic overview of the scMamba}
    \textbf{a}, The workflow of scMamba. The model employs self-supervised learning to integrate single-cell multi-omics data. 
    The core architecture of scMamba contains two encoders, each comprising stacked scMamba blocks. The scMamba block consists of alternating layers of Mamba2 and Multi-Layer Perceptron (MLP) blocks. Prior to each block, residual connection and Layer Normalization are performed.
    The encoders transform input embeddings into resulting representations,  which are subsequently passed through output embedding modules (e.g., RNA Embedding) to produce cell embeddings. These embeddings are optimized via a novel contrastive learning approach to align embeddings across omics layers.
    \textbf{b}, Detailed view of the input data embeddings. 
    Genes or chromatin accessibility peaks are first ordered according to their genomic coordinates. Subsequently, the sequencing data from each cell are partitioned into patches, each corresponding to a genomic region. These patches are linearly projected into a latent embedding space using a trainable transformation matrix. The standard learnable one-dimensional position embedding is added to retrain positional information in genomes.
    \textbf{c}, Contrastive single-cell multi-omics integration. 
    scMamba integrates single-cell multi-omics data through a novel contrastive learning approach that incorporates both a standard contrastive loss and a cosine similarity regularization term to promote cross-modal alignment of cell representations.
    }
    \label{fig:fig1}
\end{figure}

\section{Results}\label{sec2}
\subsection{Overview of scMamba}\label{subsec1}
Single-cell multi-omics sequencing technologies enable the simultaneous profiling of multiple omics layers at the individual cell level. To fully leverage the multi-omics data simultaneously profiled from single cells, we introduce scMamba, a foundation model designed to integrate single-cell multi-omics data. 
An overview of the model is depicted in Fig. \ref{fig:fig1}. 
scMamba consists of three main components: (1) patch-based cell tokenization (Fig. \ref{fig:fig1}b), (2) two scMamba encoders built from stacked scMamba blocks (Fig. \ref{fig:fig1}a), and (3) contrastive single-cell multi-omics integration (Fig. \ref{fig:fig1}c). 

To efficiently handle the high-dimensional feature space of single-cell data, scMamba introduces a patch-based cell tokenization strategy that treats genomic regions as words and cells as sentences (Fig. \ref{fig:fig1}b). 
Specifically, genes or chromatin accessibility peaks are first ordered according to their genomic coordinates.
Subsequently, the sequencing data from each cell are partitioned into patches, each corresponding to a genomic region. 
These patches are linearly projected into a latent embedding space using a trainable transformation matrix.
To retain the genomic positional information, we add the positional encoding. The standard learnable one-dimensional position embeddings are used, similar to ViT \cite{dosovitskiy2020image} and ViM \cite{zhu2024vision}.
For each modality, the input embeddings are constructed by combining patch embeddings with position embeddings.  
Next, two scMamba encoders learn to extract biologically meaningful information from each modality. Each encoder consists of stacked scMamba blocks, taking input embeddings derived from the respective modality and producing resulting representations. 
To represent cells, we need to integrate the resulting representations into a cell-embedding vector. Rather than appending a classification token\cite{devlin2019bert} to the beginning of the input tokens for cell representation, we utilize the last token of the resulting representations. This token aggregates information from input tokens, providing a comprehensive representation of the cell. 
The last token is fed into a modality-specific output embedding module, such as RNA embedding, which generates the final cell embedding. 
The final cell embeddings are optimized through a novel contrastive learning approach, incorporating an additional cosine similarity regularization module (Fig. \ref{fig:fig1}c). This design not only enhances multimodal learning but also facilitates the alignment of cell embeddings across different omics layers.
By leveraging complementary information from each modality, scMamba provides a more holistic understanding of cellular states and dynamics.

\begin{figure}[htbp]
    \centering
    \includegraphics[width=1.0\textwidth]{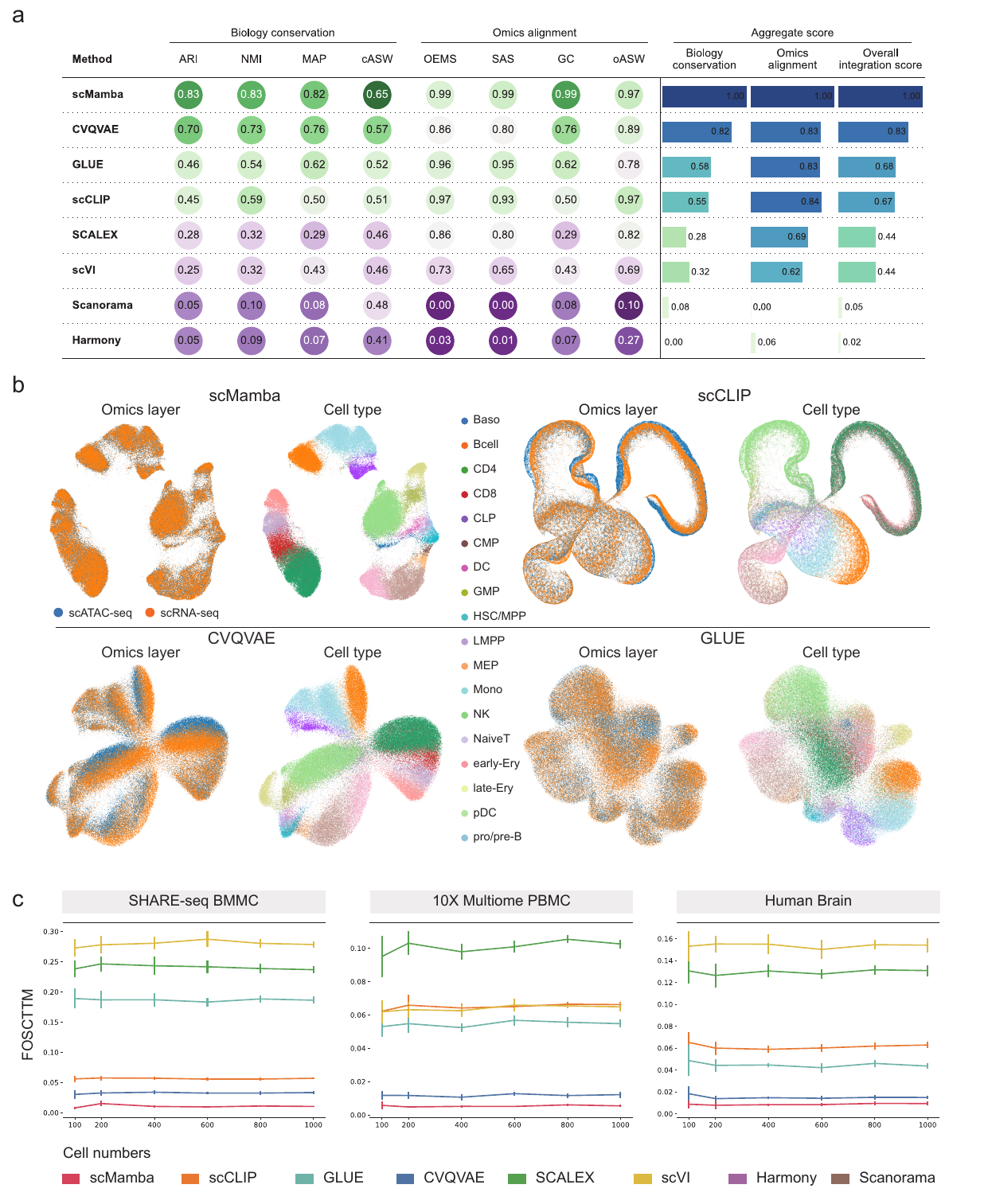}
    \caption{\textbf{Systematic benchmarks of integration performance.} \label{fig:fig2}
    \textbf{a}, The performance of scMamba and other integration methods in the SHARE-seq BMMC dataset, evaluated in terms of biological conservation, omics alignment, and aggregate scores.
    \textbf{b}, UMAP visualizations of cell embeddings in the SHARE-seq BMMC dataset after integration using the indicated methods. Cells are colored by omics layers (left) and cell types (right).
    \textbf{c}, Benchmarking single-cell level alignment error among integration methods.
    Harmony and Scanorama were excluded to avoid confounding the comparison among the remaining methods, as their markedly higher alignment errors would dominate the scale of the figure.
    }
\end{figure}

\subsection{The superior performance of scMamba in single-cell multi-omics integration}\label{subsec2}
To evaluate the integration capability of scMamba, we conducted a benchmark analysis on three scRNA-seq and scATAC-seq datasets generated by different technologies (SNARE-seq \cite{chen2019high}, SHARE-seq \cite{ma2020chromatin}, and 10X Genomics Multiome), comparing it with seven state-of-the-art integration methods \cite{xiong2023scclip, cao2022multi, liu2022cvqvae, xiong2022online, lopez2018deep, hie2019efficient, korsunsky2019fast} published recently.

An effective integration method should match the corresponding cell states from different omics layers, producing cell embeddings that preserve biological variation and align different omics layers \cite{cao2022multi, stanojevic2022computational}. To assess the performance of various methods on cell clustering and preservation of biological variation, we used four metrics, including adjusted rand index (ARI), normalized mutual information (NMI), mean average precision (MAP), and average silhouette width of cell types (cASW). Similarly, to assess the performance of various methods in omics alignment, we applied four metrics, including omics entropy mixing score (OEMS), seurat alignment score (SAS), graph connectivity (GC), and average silhouette width of omics (oASW). The overall performance was calculated by averaging these metrics after applying min-max scaling, termed the biological variation conservation and omics alignment scores. To derive the overall integration score, we applied a 6:4 weighting ratio between biological variation conservation and omics alignment, following the recommendation of a recent benchmarking study \cite{luecken2022benchmarking} as an additional criterion.

In comparison to other methods, scMamba achieved a high level of biology conservation and omics alignment simultaneously (Fig. \ref{fig:fig2}a and Supplementary Fig. 1) and was consistently the best method across all benchmark datasets in terms of aggregate score. In particular, scMamba demonstrated excellent performance in omics alignment, achieving the highest scores among all relevant metrics. In the SHARE-seq BMMC dataset (Fig. \ref{fig:fig2}a), scMamba remarkably preserved biological variation and achieved accurate clustering, whereas other methods failed to achieve comparable integration performance.

An optimal integration method should produce accurate alignments not only at the cell type level but also at the finer scale of the single-cell level. By leveraging the ground truth cell-to-cell correspondences within the datasets, we quantified single-cell alignment errors using FOSCTTM (Fraction of Samples Closer Than the True Match) \cite{singh2020unsupervised}. Additionally, we evaluated the average confidence in the accuracy of the pairings using the matching score. A lower FOSCTTM score indicates more accurate integration, as the true match is generally closer than most other potential matches. A higher matching score indicates a stronger alignment between profiles measured from the same cell across different omics layers. Across all three datasets, scMamba consistently outperformed state-of-the-art methods over a range of subsampled cell numbers, spanning from 100 to 1000 (Fig. \ref{fig:fig2}c and Supplementary Fig. 2). 
Compared to CVQVAE across all subsampling levels, scMamba achieved an approximately 90\% improvement in the matching score while consistently minimizing single-cell alignment errors.
These results highlighted the superior alignment capabilities of scMamba, demonstrating its robustness and precision in handling diverse datasets. 

To visually assess the performance of integration methods, we applied Uniform Manifold Approximation and Projection (UMAP) to visualize the integrated cell embeddings. 
The results (Fig. \ref{fig:fig2}b and Supplementary Figs. 3-5) illustrated that scMamba achieved superior alignment of shared cell types across omics layers while maintaining the distinct separation of cell populations.
By contrast, scCLIP, CVQVAE, and GLUE succeeded in aligning omics layers but encountered difficulties in accurately delineating cell-type boundaries.
Both SCALEX and scVI faced challenges in effectively aligning the omics layers and struggled to differentiate between cell types.
These results underscored the robustness of scMamba, demonstrating its capacity to achieve accurate integration while preserving biological variation across omics layers.

\begin{figure}[htbp]
    \centering
    \includegraphics[width=0.99\textwidth]{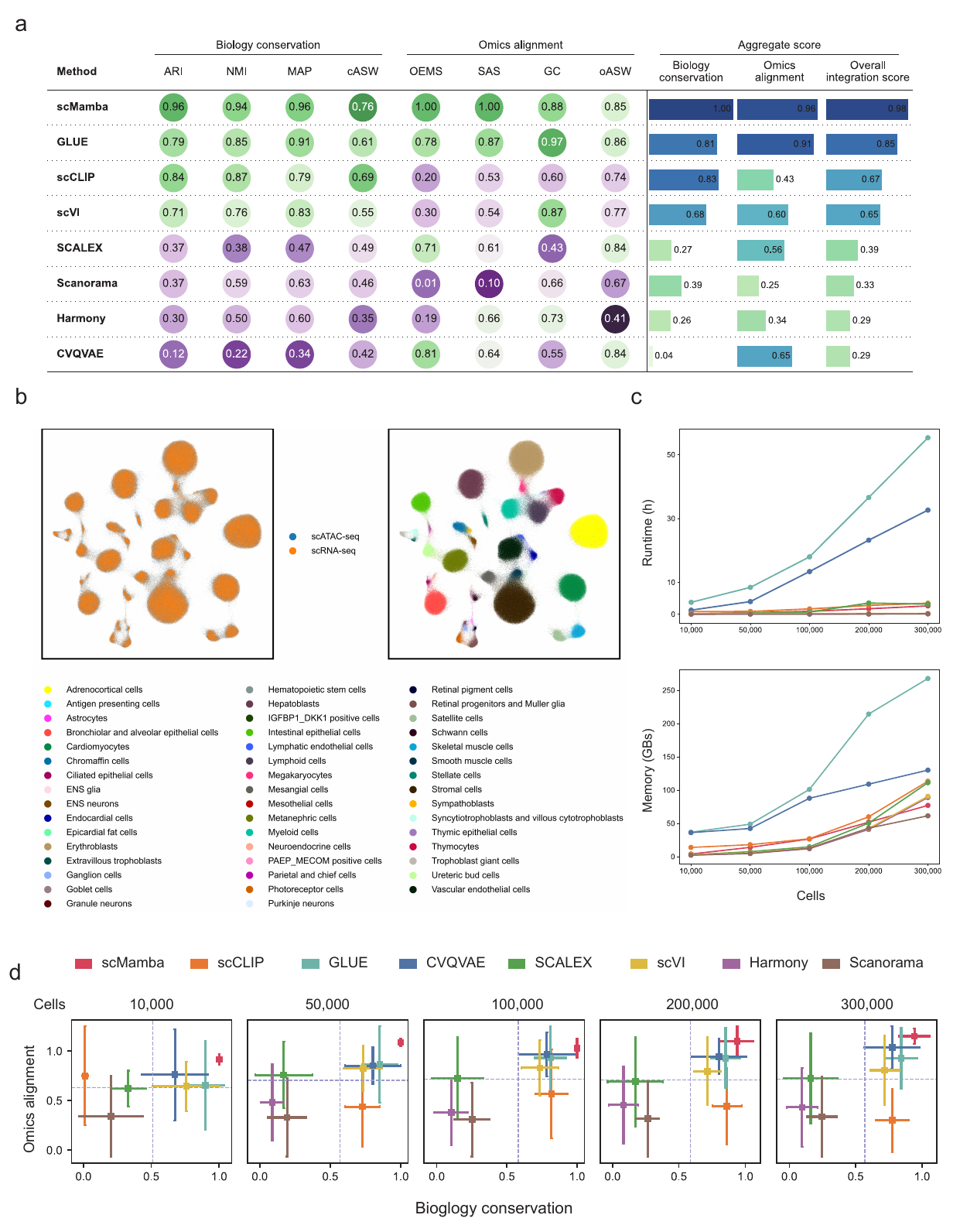}
    \caption{\textbf{Atlas-scale integration of single-cell multi-omics data.} \label{fig:fig3}
    \textbf{a}, Benchmark of the scMamba and other methods on the human fetal atlas dataset for the omics alignment and biological conservation.
    \textbf{b}, UMAP visualizations of cell embeddings in the human fetal atlas dataset after integration using scMamba, colored by omics layers (left) and cell types (right).
    \textbf{c}, Comparison of computation efficiency based on subsets of different sizes sampled from the fetal atlas dataset, including runtime (upper) and memory usage (lower). 
    \textbf{d}, Biological conservation scores versus omics alignment scores for various integration methods applied to subsets of different sizes sampled from the fetal atlas dataset. Error bars represent the standard deviation (s.d.) of the biological conservation or omics alignment metrics. Data are presented as the mean $\pm$ 0.5 times the s.d. The dashed lines are the medians of all the method results.
    }
\end{figure}

\subsection{scMamba enables atlas-scale integration}\label{subsec3}
With the continuous advancement of technology, the throughput of single-cell experiments is constantly increasing. Atlas-level single-cell multi-omics datasets that contain a large number of cells and consist of multiple tissue types have emerged \cite{luecken2022benchmarking}.
Integrating these atlas-level data poses a challenge to computational methods due to the sheer volume of data, extensive heterogeneity, and low coverage per cell \cite{cao2022multi}. 
To assess the performance of scMamba for atlas-scale integration, we applied it to the human fetal atlas dataset \cite{xiong2023scclip}, which includes RNA expression and chromatin accessibility profiles from 377,134 cells.

We evaluated the computational efficiency of scMamba and other integration methods on down-sampled subsets of the human fetal atlas dataset, ranging from 10,000 to 300,000 cells. The results revealed trends in runtime, memory usage, and integration performance as cell numbers increased (Fig. \ref{fig:fig3}c, Fig. \ref{fig:fig3}d, and Supplementary Figs. 6-8). 
Among these methods, scMamba, scCLIP, SCALEX, and scVI exhibited superior computational efficiency, with runtime and memory usage scaling nearly linearly with dataset size. However, only scMamba maintained high integration performance, while scCLIP, SCALEX, and scVI failed to achieve satisfactory integration.
Although GLUE and CVQVAE achieved satisfactory integration results, they incurred substantial computational costs. 
The performance of CVQVAE declined markedly when applied to the human fetal atlas dataset. GLUE required substantial computational resources, consuming approximately 55 hours and 268 GB of memory when applied to the 300,000-cell dataset.
scMamba achieved a unique balance of computational efficiency and integration quality, making it an ideal tool for large-scale single-cell multi-omics data integration. 

Furthermore, we discussed the influence of different numbers of highly variable genes (HVGs) and highly variable peaks (HVPs) for scMamba on multi-omics data integration (Supplementary Table 1). 
The results showed that integration performance is optimal when using all genes and peaks. Compared to 4k HVGs and 8k HVPs, selecting 8k HVGs and 16k HVPs significantly improves the conservation of biological variation.
We further evaluated the impact of varying the number of HVGs and HVPs on other methods (Supplementary Tables 2-4). 
The results indicated that increasing HVGs and HVPs enhanced the integration performance of scCLIP. However, the overall performance remained limited.
For the remaining methods, increasing HVGs and HVPs did not improve integration performance, suggesting their limited ability to effectively leverage the information embedded in high-dimensional, sparse single-cell data.
Realizing the full potential of high-dimensional, sparse single-cell data demands integration methods capable of capturing informative signals from raw features.
Unlike other methods, scMamba efficiently processes tens of thousands of genes and peaks, enabling the extraction of biologically relevant signals from raw features and leading to improved integration performance.

We successfully constructed a unified multi-omics human cell atlas by integrating RNA expression and chromatin accessibility data (Fig. \ref{fig:fig3}a and Fig. \ref{fig:fig3}b). 
Systematic benchmarking in the human fetal atlas dataset demonstrated the superior performance of scMamba (Fig. \ref{fig:fig3}a). Compared to other methods, scMamba achieved a high level of biological variation and omics alignment simultaneously and was consistently the best method in terms of aggregate score.
The human fetal atlas UMAP (Fig. \ref{fig:fig3}b) visualization not only showcased the consistent alignment of scRNA-seq and scATAC-seq embeddings achieved by scMamba but also revealed distinct separation of cell populations, with each color representing a different cell type. For comparison, we also attempted to perform integration using other methods (Supplementary Figs. 9-12). 
Although scCLIP and GLUE achieved integration at full scale, both showed notable limitations: scCLIP failed to fully align the two modalities, and GLUE could not adequately preserve biological variation after integration.
Due to its remarkable scalability and ability to accurately preserve biological variation, scMamba facilitates the integration of atlas-scale single-cell multi-omics data. This enables the comprehensive construction of cross-modal cellular atlases and provides insights into cellular heterogeneity and function.

\begin{figure}[htbp]
    \centering
    \includegraphics[width=0.89\textwidth]{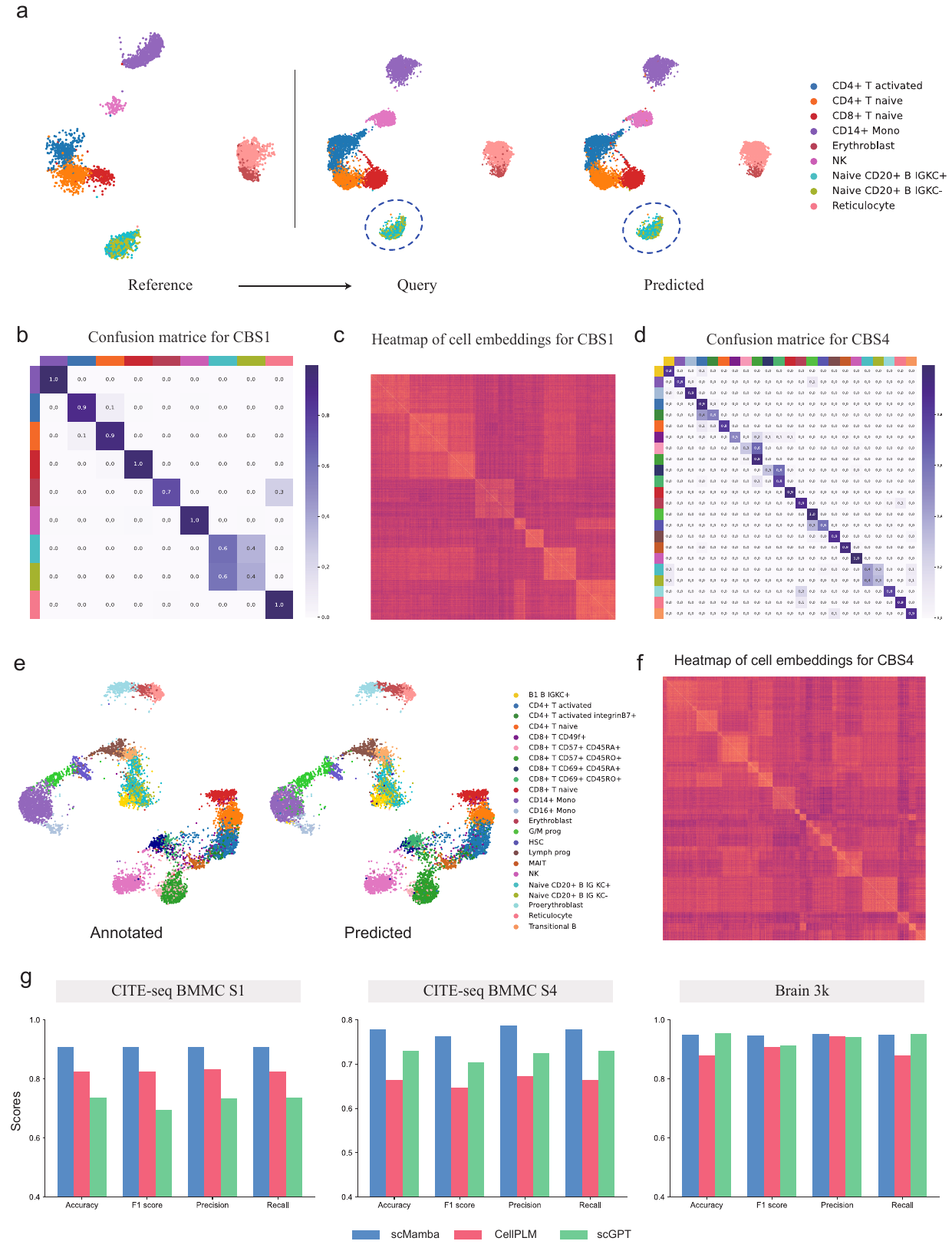}
    \caption{\textbf{Cell type annotation results using scMamba.} \label{fig:fig4}
    \textbf{a}, UMAP visualization of cell embeddings generated by scMamba from the CITE-seq BMMC S1 reference dataset, colored according to the cell types annotated (left). UMAP visualization of cell embeddings generated by scMamba from the CITE-seq BMMC S1 query dataset, colored according to the cell types annotated (middle) and the cell types predicted by scMamba (right).
    The blue circles highlight cells labeled as `Naive CD20+ B IGKC+' and `Naive CD20+ B IGKC-'.
    \textbf{b}, The confusion matrix between predicted and annotated cell types in the CITE-seq BMMC S1 dataset.
    \textbf{c}, Heatmap of cell embeddings in scMamba for cells in the CITE-seq BMMC S1 dataset, with cells grouped by annotated cell types.
    \textbf{d}, The confusion matrix between predicted and annotated cell types in the CITE-seq BMMC S4 dataset.
    \textbf{e}, UMAP visualization of cell embeddings generated by scMamba from the CITE-seq BMMC S4 dataset, colored according to the cell types annotated in the original study (left) and the cell types predicted by scMamba (right).
    \textbf{f}, Heatmap of cell embeddings in scMamba for cells in the CITE-seq BMMC s4 dataset, with cells grouped by annotated cell types.
    \textbf{g}, The barplot of accuracy, F1 score, precision, and recall compared between scMamba, CellPLM, and scGPT across three datasets.
    }
\end{figure}

\subsection{scMamba improves the precision of cell type annotation}
Cell type annotation is critical for unraveling cellular heterogeneity across tissues, developmental stages, and organisms, which in turn enhances our understanding of cellular functions in both health and disease \cite{chen2023transformer}. 
scMamba generates cell embeddings crucial for accurate cell type annotation by integrating single-cell multi-omics data.
Specifically, a neural network classifier utilizes the cell embeddings generated by scMamba as input and outputs categorical predictions for cell types. 
To evaluate the performance of scMamba for cell type annotation, we conducted extensive experiments on diverse datasets.

First, we applied scMamba to the CITE-seq BMMC (bone marrow mononuclear cells) S1 dataset for cell type prediction and visualized the results (Fig. \ref{fig:fig4}a), which incorporates RNA expression and surface protein abundance. This dataset consists of three batches, with one batch randomly selected as the reference data and the remaining two as query data.
The confusion matrix indicates that scMamba achieved high accuracy ($>$0.9) for most cell types (Fig. \ref{fig:fig4}b).
The only exceptions were for `Naive CD20+ B IGKC+' and `Naive CD20+ B IGKC-', which are subtypes of B cells with highly similar gene expression. The limited sample size in the reference partition may hinder accurate differentiation between these subtypes.
%
In addition, we visualized the cell embeddings integrated by scMamba, which demonstrate high intra-cell type similarities (Fig. \ref{fig:fig4}c).
We further tested the model on the CITE-seq BMMC S4 dataset. The reference dataset contains 6,132 cells from 23 cell types. Among them, five cell types have fewer than 100 samples, and 12 cell types have fewer than 200 samples. The limited sample size of cell types makes accurate annotation particularly challenging. 
The results showed scMamba alignment with the cell type annotations provided by the original study, achieving a high accuracy of approximately 0.78 (Fig. \ref{fig:fig4}d and Fig. \ref{fig:fig4}e).
Except for some cell subtypes that are difficult to distinguish, most cell types were correctly annotated, and the integrated cell embeddings showed clear intra-cell type consistency (Fig. \ref{fig:fig4}f). 

\begin{figure}[htbp]
    \centering
    \includegraphics[width=1.0\textwidth]{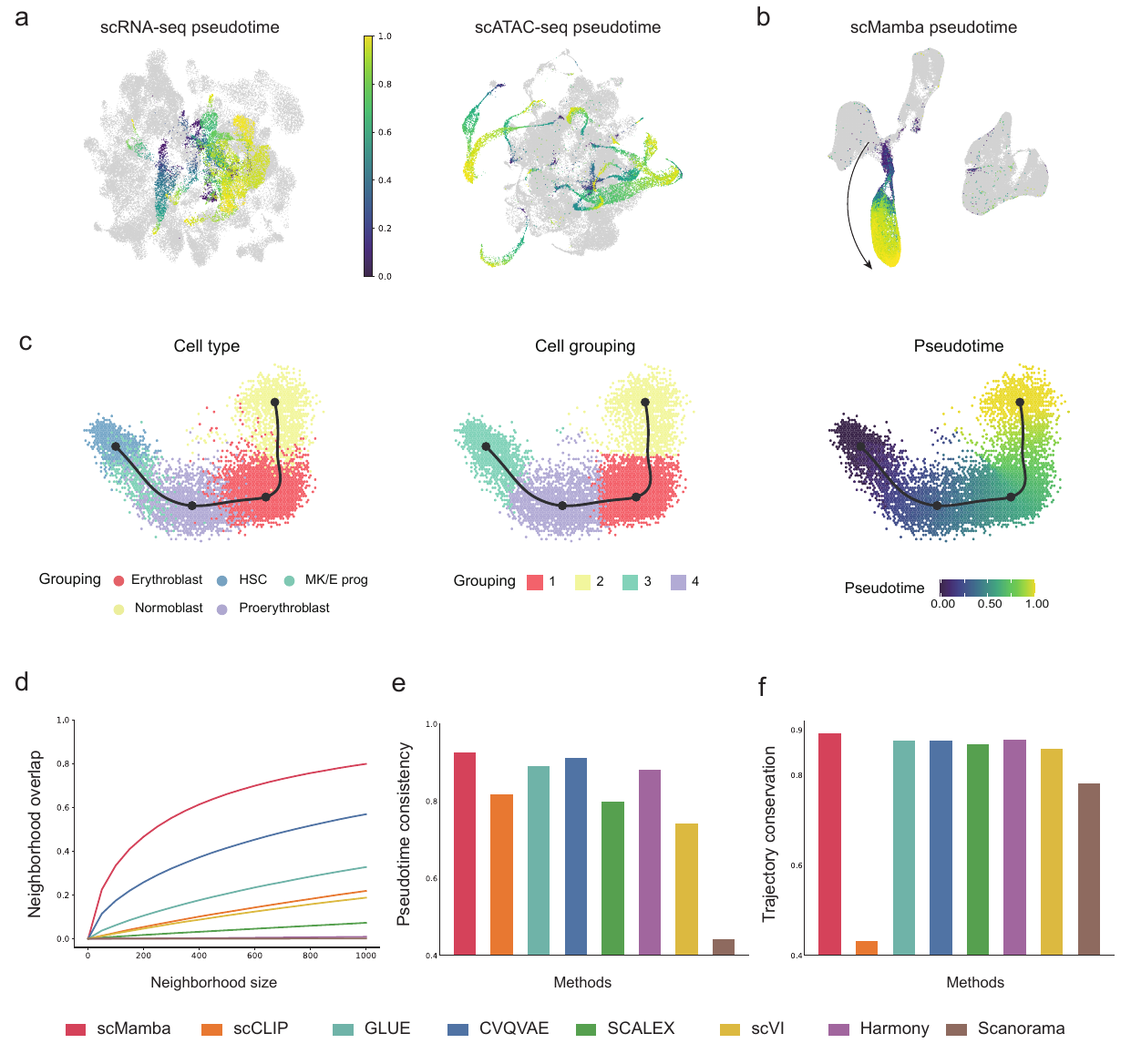}
    \caption{\textbf{Cell trajectory reconstruction results using scMamba.} \label{fig:fig5} 
    \textbf{a}, UMAP visualization of cell developmental trajectory on the original scRNA-seq data (left) and scATAC-seq data (right) from the 10x Multiome BMMC dataset, colored by cell pseudotime.
    \textbf{b}, UMAP visualizations of cell embeddings in the SHARE-seq BMMC dataset after integration using scMamba, colored by cell pseudotime. The arrow indicates inferred developmental trajectories among cell states.
    \textbf{c}, The trajectory UMAP visualization showing the original cell grouping, reconstructed groupings based on trajectory milestones, and pseudotime.
    \textbf{d}, Evaluation of integration performance using neighborhood overlap scores.
    \textbf{e}, Pseudotime consistency scores of different integration methods, evaluated using Pearson correlation.
    \textbf{f}, Trajectory conservation scores of different integration methods, averaged across scRNA-seq and scATAC-seq data.
    }
\end{figure}

Finally, we applied scMamba to a more challenging scenario, using single-cell multi-omics data as a reference to annotate single-omics data. 
In this experiment, we applied scMamba to integrate the human brain dataset and generate cell embeddings. Subsequently, we used its scRNA-seq encoder to produce cell embeddings for the human brain 3k dataset. These cell embeddings were then used for cell type prediction.
The results demonstrated that scMamba achieved high accuracy ($>$0.9) in cell type annotation (Supplementary Fig. 16a and Supplementary Fig. 16b), and the cell embeddings exhibited clear separability across different cell types (Supplementary Fig. 16c).
Finally, we benchmarked scMamba against scGPT \cite{cui2024scgpt} and CellPLM \cite{wen2023cellplm} for cell type annotation across all three datasets.
scMamba consistently outperformed both methods in key classification metrics, including accuracy, precision, recall, and F1 score, demonstrating its superiority in cell type annotation tasks (Fig. \ref{fig:fig5}g).
By effectively leveraging multiple omics layers in multimodal data, scMamba generates high-resolution cell embeddings, improving cell type annotation and advancing our understanding of complex biological processes and disease mechanisms.

\subsection{scMamba reconstructs cell differentiation trajectories}
To derive meaningful biological insights, multimodal integration approaches must effectively capture cellular state transitions, branching differentiation pathways, and gradual functional alterations. This requires the reconstruction of cellular trajectories while preserving differentiation-associated information across multiple omics layers.
To evaluate the capacity of scMamba to reconstruct cellular trajectories, we applied it to the 10x Multiome BMMC dataset consisting of RNA expression and chromatin accessibility from 69,249 human bone marrow mononuclear cells across 13 batches. 
The dataset captures well-defined stages of human erythroid differentiation, including haematopoietic stem cells (HSCs), megakaryocyte/erythroid (MK/E) progenitors, proerythroblasts, erythroblasts, and normoblasts.
%
Before integration, developmental trajectories were not discernible within either the scRNA-seq or scATAC-seq data (Fig. \ref{fig:fig5}a). After integration, scMamba reduced batch effects and aligned omics layers, resulting in a coherent and continuous trajectory aligned with known hematopoietic differentiation pathways (Fig. \ref{fig:fig5}b). 
Trajectory inference on cell embeddings generated by scMamba further resolved discrete lineage branches, delineating the progressive transitions of cell states along continuous developmental paths (Fig. \ref{fig:fig5}c).
These results indicate that scMamba accurately reconstructs cell differentiation trajectories while facilitating the visualization of dynamic cell state transitions.

Given that the 10x Multiome BMMC dataset includes ground-truth cell matching information, we utilized the neighborhood overlap score \cite{korsunsky2019fast, cao2020unsupervised} to benchmark the integration performance of scMamba and other methods. This metric evaluates the alignment of nearest neighbors between omics layers across varying neighborhood sizes. 
The result showed that scMamba consistently achieved the highest neighborhood overlap scores across a wide range of sample sizes, from 1 to 1000 cells, demonstrating superior omics matching performance (Fig. \ref{fig:fig5}d).

To further assess the preservation of developmental structure, we examined pseudotime consistency between matched cells across omics layers.  
Pseudotime was inferred independently for each omics layer in the latent space, and cross-omics consistency was evaluated using Pearson correlation between matched pseudotime values. 
scMamba exhibited the highest consistency across both RNA and ATAC layers, with CVQVAE also performing well (Fig. \ref{fig:fig5}e). 
Finally, we calculated the trajectory conservation scores to quantify how well each method retained differentiation patterns within each omics layer.
scMamba again outperformed all baselines, achieving the highest trajectory conservation scores in both scRNA-seq and scATAC-seq data  ( Fig. \ref{fig:fig5}f). 
These findings confirm that scMamba effectively preserves cell differentiation trajectories during single-cell multi-omics integration, providing a framework for trajectory analysis across diverse omics layers.

\section{Discussion}\label{sec3}
In this study, we introduce scMamba, a foundation model designed to integrate single-cell multi-omics data, including RNA expression, chromatin accessibility, and surface protein abundance. 
Unlike existing methods, which often suffer from information loss due to feature selection or dimensionality reduction during preprocessing, scMamba directly processes the raw features without the selection of genes or peaks, enabling the extraction of more comprehensive and biologically relevant information. 
Ablation experiments showed that preserving raw features significantly enhances integration performance, underscoring the importance of maintaining full-resolution biological information for downstream analyses.
The scMamba encoder leverages the state space duality algorithm to extract meaningful biological insights from raw single-cell features, enabling accurate characterization of cell identity and differentiation trajectories.

Since most single-cell multi-omics data are unlabeled, we designed a self-supervised training strategy based on a novel contrastive learning approach, incorporating an additional cosine similarity regularization module. 
This approach enhances the multimodal learning of biological signals, while cosine similarity regularization further aligns cell embeddings across omics layers, leading to improved integration and clustering accuracy.
Existing single-cell foundation models \cite{cui2024scgpt, yang2022scbert, hao2024large} treat genes as words and cells as sentences. However, these approaches are limited to processing around 2,000 genes due to computational and architectural limitations, thereby restricting their applicability to more complex and comprehensive genomic landscapes.
To overcome this limitation, scMamba introduces an innovative patch-based cell tokenization strategy that treats genomic regions as words and cells as sentences. Genomic positional information is preserved through positional encoding. 
This strategy enhances computational efficiency while preserving critical genomic context, facilitating more expressive and scalable single-cell analysis.

Systematic benchmarking demonstrates that scMamba consistently outperforms existing state-of-the-art methods in integrating single-cell multi-omics data.
scMamba accurately integrates multimodal data at the single-cell level while preserving biologically meaningful heterogeneity, thereby enabling high-resolution mapping of cellular states and reconstruction of intricate differentiation pathways.
As the availability of single-cell multi-omics data continues to expand, we anticipate that scMamba will serve as a foundational tool for uncovering cellular diversity and dynamics in diverse biological contexts, ranging from tissue development to disease progression and therapeutic response.
Beyond multimodal integration, scMamba provides a scalable and generalizable framework for modeling single-cell data across modalities. Its raw feature-level processing and patch-based cell tokenization strategy facilitate efficient and expressive representation learning, making it particularly well-suited for large-scale and heterogeneous data.
                                                               

\section{Methods}\label{sec4}
\subsection{scMamba}\label{subsec1}
\subsubsection{Input embedding}\label{subsubsec1}
Single-cell multi-omics sequencing technologies enable the simultaneous profiling of RNA expression (scRNA-seq) and chromatin accessibility (scATAC-seq) or RNA expression and surface protein abundance at the individual cell level. RNA expression data are processed into a cell-by-gene matrix, $X^{R}\in \mathbb{R}^{N\times G}$, where each element ${X}_{ij}^{R}\in \mathbb{R}^{+} $ indicates the RNA abundance for gene $j$ in cell $i$. Chromatin accessibility data are represented as a cell-by-peak matrix, $X^{A}\in \mathbb{R}^{N\times P}$, with $X_{ij}^{A}\in \mathbb{R}^{+}$ denoting the chromatin accessibility at peak region $j$ in cell $i$. Additionally, surface protein abundance data are represented as a cell-by-protein matrix, $X^S \in \mathbb{R}^{N\times Q}$, where $X_{ij}^{S} \in \mathbb{R}^{+}$ indicates the abundance of surface protein $j$ in cell $i$. 
These matrices will be referred to as the raw count matrices in the following sections.

\noindent\textbf{Single-cell data preprocessing.}
For scRNA-seq data $X^{R}$, we normalize RNA expression by the total read count per cell and apply a log1p transformation \cite{stuart2019comprehensive}.
For scATAC-seq data $X^{A}$, we binarize peak accessibility, assigning a value of 1 to peaks with nonzero counts and 0 otherwise. Surface protein abundance data $X^{S}$ are used without preprocessing.

\noindent\textbf{Expression value encoding.}
Variability in absolute expression magnitudes across sequencing protocols poses a challenge for integrating RNA expression and surface protein abundance data. Differences in sequencing depth and gene sparsity introduce batch-specific scale variations that normalization cannot fully resolve. Thus, identical expression values may reflect distinct biological contexts across batches. To address this, we bin expression values based on relative levels across cells \cite{yang2022scbert, cui2024scgpt}, partitioning nonzero counts into $B$ consecutive intervals $[b_k, b_{k+1}]$, ensuring consistency in semantic interpretation across datasets. 
It is important to note that the scATAC-seq cell-by-peak matrix is not binned but rather binarized, encoding chromatin accessibility as presence or absence.

\noindent\textbf{Cell Tokenization.}
Tokenization is a crucial step in processing high-dimensional input data and has been fundamental to the success of Mamba-based and Transformer-based models \cite{chen2023transformer, gu2023mamba, zhao2024cobra, huang2024ml}. 
Inspired by approaches used in image processing \cite{alexey2020image, zhu2024vision}, we adapt tokenization specifically for single-cell profiles, which treats genomic regions as words and cells as sentences. Each genomic region represents the set of genes or chromatin accessibility peaks located within it, ordered according to their genomic coordinates. 
Specifically, genes (or chromatin accessibility peaks) are first ordered based on their locations on the genome. 
For each cell, the sequencing profile vector $x \in \mathbb{R}^{L}$ is then partitioned into patches $x_p \in \mathbb{R}^{C\times P}$, where $L$ is the length of the profile, $P$ is the patch size, and $C$ is the number of patches. 
The patches are linearly projected into vectors of size $D$ using a trainable transformation matrix.
Zero-padding is applied to standardize token lengths. 
Position embeddings are added to the patch embeddings to retrain positional information in genomes. We use standard learnable one-dimensional position embeddings $E_{pos} \in \mathbb{R}^{C\times D}$. 
Consequently, the final embedding $T^{(i)}\in \mathbb{R}^{C\times D}$ for cell $i$ is defined as: 
\begin{equation}  \label{equ:equ2}
    T = [t^{(1)}W;t^{(2)}W;\cdots;t^{(C)}W] + E_{pos}.
\end{equation}
Here, $t^{(i)}$ denotes the $i$-th patch for single-cell profiles, while $C$ represents the number of patches. The learnable projection matrices $W\in \mathbb{R}^{P\times D}$ transform the patches to the desired embedding dimensions $D$.

\subsubsection{Model architecture}\label{subsubsec2}
The architecture of scMamba is illustrated in Fig. \ref{fig:fig1}. The framework incorporates two modality-specific encoders designed to extract meaningful features from scRNA-seq and scATAC-seq data. The encoder is composed of a stack of $L$ identical scMamba blocks. 
The scMamba block consists of alternating layers of Mamba2 and MLP blocks. Before each block, residual connection and Layer Normalization are performed.
Distinct from the standard Transformer architecture \cite{vaswani2017attention}, scMamba introduces residual connections and layer normalization before the Mamba2 and MLP blocks, rather than following them. 
This design improves computational efficiency by enabling adaptive input feature scaling, thereby enhancing model performance in single-cell analyses.
We use the scMamba to encode the input embedding $T^{(i)}$ in equation (\ref{equ:equ2}). The state space duality (SSD) algorithm of Mamba2 operates on the sequence of $C$ embedding vectors, extracting biologically meaningful insights from raw genomic features for cell representation.
The output of the stacked scMamba blocks can be defined as follows:
\begin{align}
    T_{0}^{i} &= T^{i} \\
    T_{l}^{i} &= \text{scmamba\_block}(T_{l-1}^{i}) \ \  \forall l \in[1, L]  
\end{align}
The resulting representation $T_{L}^{(i)}\in \mathbb{R}^{C\times D}$ is used for cell-level tasks. Specifically, we first integrate $T_{L}^{(i)}$ into a cell-embedding vector (Cell representation). 
Leveraging the SSD algorithm, scMamba extracts critical biological insights from high-dimensional, sparse single-cell sequencing data while achieving an optimal balance across key efficiency metrics, including training and inference computation, memory usage, and efficient utilization of matrix multiplication units on modern hardware \cite{dao2024transformers}.
The SSD algorithm simplifies the state transition matrix $A$ to a scalar $a$ multiplied by the identity matrix, thereby reducing computational complexity while retaining expressiveness.
\begin{equation}
    L=\operatorname{1SS}\left(a_{0: C}\right):=\left[\begin{array}{ccccc}
    1 & & & & \\
    a_1 & 1 & & & \\
    a_2 a_1 & a_2 & 1 & & \\
    \vdots & \vdots & \ddots & \ddots & \\
    a_{C-1} \ldots a_1 & a_{C-1} \ldots a_2 & \ldots & a_{C-1} & 1
    \end{array}\right]
\end{equation}
Here, 1SS denotes 1-Semiseparable matrices \cite{dao2024transformers}, a simplified state-space structure in which the state transition matrix is reduced to a scalar multiplied by the identity matrix.
The SSD algorithm forms a direct connection between structured state-space representations and attention mechanisms:
\begin{align}
    M & = L \circ\left(C B^{\top}\right) \\
    y & = Mx
\end{align}
This approach differs from standard softmax attention in two key respects: the omission of the softmax function and the elementwise multiplication of the attention matrix by an additional mask matrix $L$.
Omitting the softmax function addresses issues observed in traditional attention, such as the ``attention sink" phenomenon \cite{xiao2023efficient, darcet2023vision}, thereby allowing scMamba to consider all relevant information instead of concentrating on a limited subset.
More significantly, the mask matrix $L$ assigns variable weights to different parts of the input, enabling scMamba to prioritize the most relevant components.

\subsubsection{Cell representation}\label{subsubsec3}
Each cell is partitioned into patches, and its representation $T_{c}^{(i)}\in \mathbb{R}^{D}$ is obtained by aggregating the learned $T_{L}^{i}$. 
Several pooling operations can be readily applied in this study, such as element-wise mean pooling \cite{lin2013network}, weighted pooling, or the introduction of a special [CLS] token for pooling \cite{devlin2019bert}.
As detailed in the model architecture section, $L$ is the causal mask that restricts each token to attend only to its preceding tokens. Under this constraint, only past information is available when generating the final token. Therefore, we use the last token as the summary representation of the cell, which is then passed through an MLP head to produce the final cell embedding.

\subsubsection{Multi-omics embedding learning}\label{subsubsec4}
Inspired by recent advances in multimodal image-text representation learning approaches \cite{radford2021learning, yu2022coca, jia2021scaling}, we propose a self-supervised training strategy based on a novel contrastive learning approach to enhance multimodal learning of cell embeddings across multiple omics layers. The approach comprises a contrastive loss and an additional cosine similarity regularization loss.
We adapt contrastive learning specifically for single-cell multi-omics data. RNA and ATAC profiles originating from the same cell are designated as positive pairs. In contrast, all other RNA-ATAC combinations randomly drawn within a batch are considered negative pairs to encourage the model to distinguish true biological correspondences.
The objective is to maximize the cosine similarity between the RNA and ATAC embeddings of the $N$ true pairs while simultaneously minimizing the cosine similarity for the $N^2 - N$ incorrect pairings. To achieve this, we minimize the sum of two losses: one for RNA-to-ATAC classification and the other for ATAC-to-RNA classification. The corresponding loss function is defined as follows:

\begin{equation}
    \mathcal{L}_{con}=-\frac{1}{N}\left(\sum_i^N \log \frac{\exp \left(x_i^{\top} y_i / \sigma\right)}{\sum_{j=1}^N \exp \left(x_i^{\top} y_j / \sigma\right)}+\sum_i^N \log \frac{\exp \left(y_i^{\top} x_i / \sigma\right)}{\sum_{j=1}^N \exp \left(y_i^{\top} x_j / \sigma\right)}\right)
\end{equation}
Here, $x_i$ and $y_i$ represent the embeddings of RNA and ATAC in the $i$-th pair, respectively, and $N$ is the batch size. The parameter $\sigma$ is the temperature to scale the logits. 
To further align the cell embeddings across the modalities, we introduce an additional cosine similarity regularization module, given by:
\begin{equation}
    \mathcal{L}_{sim} = 1 - \frac{1}{N}\sum_{i}^{N}{\cos(x_i, y_i)}
\end{equation}
where $\cos(x_i, y_i)$ denotes the cosine similarity between the RNA and ATAC embeddings for the $i$-th pair. 
This module promotes the alignment of cell embeddings across modalities, thereby enabling a unified and biologically meaningful representation of the multi-omics data.

The overall training objective combines both components:
\begin{equation}
    \mathcal{L} = \lambda_{con} \cdot \mathcal{L}_{con} + \lambda_{sim} \cdot \mathcal{L}_{sim}
\end{equation}
where $\lambda_{con}$ and $\lambda_{sim}$ balance the relative contribution of contrastive loss and cosine similarity regularization. 
This formulation facilitates both the discriminative separation of mismatched pairs and the fine-grained alignment of matched cells, enabling robust multimodal learning across scRNA-seq and scATAC-seq modalities.

\subsection{Systematic benchmark}
\subsubsection{Evaluation metrics}
\noindent\textbf{ARI and NMI}. 
To evaluate the agreement between known cell-type labels and clusters generated by the Leiden algorithm, we employed the Adjusted Rand Index (ARI) and Normalized Mutual Information (NMI) as key metrics. To mitigate potential biases stemming from the resolution parameter of the Leiden algorithm, we performed clustering across a resolution range of 0.1 to 2.0 in 0.1 increments, using integrated results from each method. The performance of each method was then determined by the highest ARI and NMI values observed across these resolutions.

\noindent\textbf{Mean average precision (MAP)}. 
The MAP is used to evaluate the cell type resolution \cite{cao2022multi}. Let the cell type of the $i$-th cell be denoted as $y_i$ and that the cell types of its $K$ nearest neighbors, ordered by proximity, as $y_{i1}$, $y_{i2}$, ..., $y_{iK}$. The mean average precision is then defined as follows:
\begin{equation}
    \text{MAP} = \frac{1}{N} \sum_{i=1}^{N}{AP_i}
\end{equation}
\begin{equation}
\text{AP}^{\left( i \right)}=\begin{cases}
	\frac{\sum_{k=1}^K{I_{y_i=y_{ik}}}\cdot \frac{\sum_{j=1}^k{y_i}=y_{ij}}{k}}{\sum_{k=1}^K{I_{y_i=y_{ik}}}},&		\,\,\text{if\,\,}\sum_{k=1}^K{I_{y_i=y_{ik}}}>0\\
	0,&		\,\,\text{otherwise\,\,}\\
\end{cases}
\end{equation}
Here, $I_{y_i=y_{ik}}$ is an indicator function that equals 1 if $y_i = y_{ik}$ and 0 otherwise. The average precision (AP) for each cell is calculated by averaging the precision for the neighbors that match the cell type. The MAP is then computed as the mean of the AP values across all cells. In our analysis, $K$ is set to 1\% of the total number of cells in each dataset. The MAP ranges from 0 to 1, with higher values indicating better resolution of cell types.

\noindent\textbf{The average silhouette width (ASW)} \cite{rousseeuw1987silhouettes}.
The ASW metric is utilized to assess the precision of cell-cell distances calculated by each integration algorithm. The ASW reflects both intra-cluster similarity (how well a cell matches other cells within its cluster) and inter-cluster dissimilarity (how distinct it is from cells in the nearest different cluster). By averaging the silhouette widths of a set of cells, the ASW is derived, with values ranging from -1 to 1.
In our analysis, we calculated ASW scores based on cell type labels (cASW) and omics layer labels (oASW) to evaluate each algorithm's ability to preserve biological variation and align omics layers, respectively. A higher cASW indicates better separation of cell types, whereas a lower oASW suggests more effective omics alignment. To standardize evaluation, we applied linear transformations to both cASW and oASW values \cite{luecken2022benchmarking}, ensuring that higher values consistently represent better performance.
\begin{equation}
    \text{cASW} = \frac{\text{ASW}_c + 1}{2}
\end{equation}
where $\text{ASW}_c$ indicates the average silhouette width grouped by cell type.

\begin{align}
    \text {o} \mathrm{ASW}_{j} &=\frac{1}{N_j} \sum_{i=1}^{N_j} 1-\left|s_{\text{o}}^{(i)}\right| \\
    \text {oASW }&=\frac{1}{M} \sum_{j=1}^M \text{o}\mathrm{ASW}_{j}
\end{align}
where $s_{\text{o}}^{(i)}$ is the omics layer silhouette width for the $i$-th cell, $N_j$ is the number of cells in cell type $j$, and $M$ is the total number of cell types. The oASW ranges from 0 to 1, and the higher values indicate better omics alignment.


\noindent\textbf{Omics entropy mixing score (OEMS)}. The OEMS is adapted from "batch entropy mixing score" \cite{xiong2022online, haghverdi2018batch} and designed to assess the regional mixing of cells from different omics layers. A high score indicates that the omics layers are well aligned. The omics entropy mixing score was calculated as follows:

(1) Calculate the proportion $P_i$ of cell numbers in each omics layer to the total cell numbers.

(2) Randomly select 100 cells from all omics layers.

(3) Calculate the 100 nearest neighbors for each randomly selected cell.

(4) The regional mixing entropies for each cell are as follows:
\begin{equation}
    p_i{'} = \frac{\frac{p_i}{P_i}}{\sum_{i=1}^{n}{\frac{p_i}{P_i}}}
\end{equation}
\begin{equation}
    \text{En} = \sum_{i=0}^{n}{p_i{'}\log{p_i{'}}}
\end{equation}
where $p_i$ represents the porportion of cells from omics layer $i$ in a given region, satisfying $\sum_{i=0}^{n}{p_i}=1$. The term $p_i'$ is a correction factor that accounts for discrepancies caused by differences in cell numbers across omics layers. The total mixing entropy is computed as the sum of regional mixing entropies.

(5) Repeat step (2)-(4) for 100 iterations with different randomly selected cells. The omics entropy mixing score is calculated as the average $\text{En}$ across all iterations.

\noindent\textbf{Seurat alignment score (SAS)}. The SAS is used to evaluate the degree of mixing among omics layers, as described in the original publication \cite{butler2018integrating}.
\begin{equation}
    \text{SAS} = 1 - \frac{\bar x - \frac{K}{N}}{K - \frac{K}{N}}
\end{equation}
where $\bar x$ represents the average number of cells from the $K$-nearest neighbors. Here, $N$ denotes the total number of omics layers, and $K$ is set to 1\% of the subsampled cell count. The SAS ranges from 0 to 1, with higher values indicating greater omics alignment.

\noindent\textbf{Graph connectivity (GC)} \cite{luecken2022benchmarking}. The GC score assesses whether the kNN graph representation $G$ of the integrated data directly connects all cells with the same cell identity label. For each cell identity label $c$, we construct a subset KNN graph $G(N_c; E_c)$, which includes only cells from a given cell type. The GC score is calculated as follows:
\begin{equation}
    \text{GC} = \frac{1}{|C|} \sum_{c\in C}{\frac{|\text{LCC}(G(N_c;E_c))|}{|N_c|}}
\end{equation}
Here, $C$ represents the set of cell identity labels, $|\text{LCC()}|$ is the number of nodes in the largest connected component of the graph, and $|N_c|$ is the number of nodes with cell identity $c$. The GC score has a $(0, 1]$ range, where 1 indicates that all cells of the same cell identity are connected in the integrated KNN graph.

\noindent\textbf{Overall integration score}. The overall integration score $S_{\text{overall}}$ is used to comprehensively assess the integration performance, which aggregates the biology conservation score $S_{bio}$ and omics alignment score $S_{omics}$. The $S_{\text{overall}}$ is calculated as follows:
\begin{equation}
    S_\text{overall} = S_\text{bio} \times 0.6 + S_\text{omics} \times 0.4
\end{equation}
The partial scores $S_{\text{bio}}$ and $S_{\text{omics}}$ are computed by averaging the scaled values of all metrics associated with each score. Specifically:
\begin{equation}
    S_{\text{bio}} = \frac{1}{|M_{\text{bio}|}}\sum_{m_i \in M_{\text{bio}}}{\text{Scale}(m_i)} 
\end{equation}
\begin{equation}
    S_{\text{omics}} = \frac{1}{|M_{\text{omics}|}}\sum_{m_i \in M_{\text{omics}}}{\text{Scale}(m_i)} 
\end{equation}
Here, $M_{\text{bio}}$ contains the metrics ARI, NMI, MAP, and cASW, representing the sets of metrics contributing to the biology conservation scores. Similarly, $M_{\text{bio}}$ contains the metrics OEMS, SAS, GC, and oASW, representing the sets of metrics contributing to the omics alignment scores.
To ensure that each metric is equally weighted within a partial score and possesses the same discriminative power, we applied min–max scaling to each metric using the function $\text{Scale()}$, defined as:
\begin{equation}
    \text{Scale}(x) = \frac{X - \text{min}(X)}{\text{max}(X) - \text{min}(X)}
\end{equation}

\noindent\textbf{FOSCTTM}. The Fraction of Samples Closer than True Match (FOSCTTM) \cite{singh2020unsupervised} is used to assess single-cell alignment errors across two omics layers with known ground truth pairings. Assuming each omics layer contains $N$ cells and the cells are sorted in the same order, we define $x$ and $y$ as the cell embeddings for the first and second omics layers, respectively.  The FOSCTTM is calculated as:
\begin{equation}
        \text { FOSCTTM }=\frac{1}{2 \mathrm{~N}}\left(\sum_{i=1}^{\mathrm{N}} \frac{n_1^{(i)}}{\mathrm{N}}+\sum_{i=1}^{\mathrm{N}} \frac{n_2^{(i)}}{\mathrm{N}}\right)
\end{equation}
where $n_1^{(i)}$ and $n_2^{(i)}$ represent the number of cells that are closer to the $i$-th cell in the opposite omics layer than its true match. Here, $d$ denotes the Euclidean distance between cell embeddings. FOSCTTM has a range of 0 to 1, and lower values indicate higher accuracy.
The values of $n_1^{(i)}$ and $n_2^{(i)}$ are define as:
\begin{equation}
    \begin{aligned}
        n_1^{(i)} &=\left|\left\{j \mid d\left(\mathbf{x}_j, \mathbf{y}_i\right)<d\left(\mathbf{x}_i, \mathbf{y}_i\right)\right\}\right| \\
        n_2^{(i)} &=\left|\left\{j \mid d\left(\mathbf{x}_i, \mathbf{y}_j\right)<d\left(\mathbf{x}_i, \mathbf{y}_i\right)\right\}\right|
    \end{aligned}
\end{equation}

\noindent\textbf{Matching Socre (MS)}. The average confidence in the correct pairing of cells across modalities is quantified using a cross-modality matching matrix, $M_{ij}$, which is constructed by calculating the Jaccard index between the nearest neighbors in the latent embedding space for each modality. Specifically, the matching matrix is defined as:
\begin{equation}
    M_{ij} = \frac{\lvert (K_{ij} \cap K_{jj}) \cup (K_{ji} \cap K_{ii}) \rvert}{\lvert (K_{ij} \cup K_{jj}) \cup (K_{ji} \cup K_{ii}) \rvert}
\end{equation}
where $K_{ij}$ denotes the nearest neighbor of the sample $i$ in the modality of a sample $j$.
The Matching Score is then calculated by normalizing the matching matrix $M$ on a per-row basis to obtain $\tilde{M}$. The MS is defined as:
\begin{equation}
    \text { Matching Score }=\frac{1}{\mathrm{~N}} \sum_i \sum_j \tilde{\mathrm{M}}_{i, j} * I_{i,j}
\end{equation}
where $I_{ij}$ equals 1 if profile $i$ and $j$ originate from the same cell, and 0 otherwise. $N$ represents the total number of observations.

\noindent\textbf{Trajectory conservation (TC)} \cite{luecken2022benchmarking}. 
The TC score assesses the preservation of cell state information following data integration. 
To compute this score, we first derive the trajectories using Scanpy on our integrated embeddings and then compare these trajectories with those from the original dataset. We calculate Spearman's rank correlation coefficient $s$ between the two sets of trajectories, which is subsequently normalized to yield the final TC score, defined as:
\begin{equation}
    \text{TC}(s) = \frac{s + 1}{2} 
\end{equation}
A higher TC score reflects better preservation of the biological information.

\noindent\textbf{Neighborhood overlap score} \cite{korsunsky2019fast, cao2020unsupervised}. 
The Neighborhood overlap score quantifies the degree of integration between datasets when cell-cell correspondence exists across different omics layers. For a given neighborhood size $k$, this score constructs a k-nearest neighbor graph based on the latent embeddings of cells derived from both scRNA-seq and scATAC-seq data. It then calculates the proportion of cells whose corresponding counterparts from the other modality are present within their k-NN neighborhood.

\subsubsection{Baseline methods}
We compared scMamba with seven single-cell multi-omics integration methods, including scCLP \cite{xiong2023scclip}, CVAVAE \cite{liu2022cvqvae}, GLUE \cite{cao2022multi}, SCALEX \cite{xiong2022online}, scVI \cite{lopez2018deep}, Harmony \cite{korsunsky2019fast}, and Scanorama \cite{hie2019efficient}.

\noindent\textbf{scCLIP}. We applied scCLIP following the implementation provided in its GitHub repository (\href{https://github.com/jsxlei/scCLIP}{https://github.com/jsxlei/scCLIP}). Following the default pipeline, the data were preprocessed into the required input format, and multimodal integration was conducted using the default parameters.

\noindent\textbf{CVQVAE}. We applied CVAVAE following the implementation provided in its GitHub repository (\href{https://github.com/HelloWorldLTY/CVQVAE}{https://github.com/HelloWorldLTY/CVQVAE}). Following the default pipeline, we normalized each cell to 10,000 reads and log-transformed the data. The top 3,000 highly variable genes were selected, and dimensionality reduction was performed using PCA for scRNA-seq and LSI for scATAC-seq, reducing the data to 2,500 dimensions. All other parameters remained at default values.

\noindent\textbf{GLUE}. We used the scglue package (version 0.3.2) provided in its GitHub repository (\href{https://github.com/gao-lab/GLUE}{https://github.com/gao-lab/GLUE}). Following its default pipeline, we selected 2,000 highly variable genes, normalized each cell to 10,000 reads, log-transformed the data, and standardized it. Dimensionality reduction was then performed using PCA for scRNA-seq and LSI for scATAC-seq, reducing the data to 100 dimensions. Finally, we constructed the prior regulatory graph. All other parameters remained at default values.

\noindent\textbf{SCALEX}. We applied CVAVAE following the implementation provided in its GitHub repository (\href{https://github.com/jsxlei/SCALEX}{https://github.com/jsxlei/SCALEX}). The scATAC-seq data are transformed into the gene activity matrices using the Python package EpiScanpy to quantify the activity of each gene. These matrices were then integrated with scRNA-seq data, treating them as separate "batches". After that, the data were preprocessed into the required input format, and multimodal integration was performed using the default parameters.

\noindent\textbf{scVI}. We used the Python package scvi-tools (version 1.3.0) provided in its GitHub repository (\href{https://github.com/scverse/scvi-tools}{https://github.com/scverse/scvi-tools}) and followed the suggested pipelines. The scATAC-seq data are transformed into the gene activity matrices using the Python package EpiScanpy to quantify the activity of each gene. We then combined the gene activity score matrix with the scRNA-seq data matrix as two individual "batches" for integration. After that, the data were preprocessed into the required input format, and multimodal integration was performed using the default parameters.

\noindent\textbf{Harmony}. We used the Python package harmonypy (version 0.0.10) provided in its GitHub repository (\href{https://github.com/slowkow/harmonypy}{https://github.com/slowkow/harmonypy}) and followed the recommended pipelines. The scATAC-seq data are transformed into the gene activity matrices using the Python package EpiScanpy to quantify the activity of each gene. These matrices were then integrated with scRNA-seq data, treating them as separate "batches". After that, the data were preprocessed into the required input format, and multimodal integration was performed using the default parameters.

\noindent\textbf{Scanorama}. We applied Scanorama following the implementation provided in its GitHub repository (\href{https://github.com/brianhie/scanorama}{https://github.com/brianhie/scanorama}) and followed the suggested pipelines. The scATAC-seq data are transformed into the gene activity matrices using the Python package EpiScanpy to quantify the activity of each gene. These matrices were then integrated with scRNA-seq data, treating them as separate "batches". After that, the data were preprocessed into the required input format, and multimodal integration was performed using the default parameters. 

\subsection{Datasets}
\noindent\textbf{SHARE-seq BMMC}. The SHARE-seq BMMC dataset contains simultaneous profiling of RNA expression and chromatin accessibility from human BMMC cells, processed using SHARE-seq alignment v2 pipeline \cite{ma2020chromatin}. The dataset includes 78,520 cells with read counts from 51,862 genes and 173,026 chromatin regions, available from NCBI GEO under accession number GSE207308.

\noindent\textbf{Brain}. The human brain dataset contains simultaneous profiling of RNA expression and chromatin accessibility from brain tissues of donors with Alzheimer’s disease (AD) and unaffected controls, sequenced by the 10x Genomics Multiome protocol. It includes 105,332 cells with read counts from 36,601 genes and 140,614 chromatin regions, downloaded from NCBI GEO under the accession number GSE214637 \cite{anderson2023single}.

\noindent\textbf{10x Multiome PBMC}. The 10x Multiome PBMC dataset contains simultaneous profiling of RNA expression and chromatin accessibility from human PBMC cells sequenced by the 10x Genomics Multiome protocol. All samples in this dataset were obtained from the same healthy donor. It includes 9,631 cells with read counts for 29,095 genes and 107,194 chromatin regions.

\noindent\textbf{Fetal Atlas}. The human fetal atlas dataset contains simultaneous profiling of RNA expression and chromatin accessibility from human fetal cells. The dataset encompasses 377,134 cells with read counts for 36,601 genes and 1,154,464 chromatin regions, providing a critical resource for understanding the molecular events that occur during human fetal development.

\noindent\textbf{CITE-seq BMMC S1}. The CITE-seq BMMC S1 dataset contains simultaneous profiling of RNA expression and protein abundance from human BMMC cells sequenced by the CITE-seq protocol \cite{stoeckius2017simultaneous}. These cells came from 12 healthy human donors consisting of 3 batches in this dataset. The dataset encompasses 11,126 cells with measurements from 13,953 genes and 134 surface proteins.

\noindent\textbf{CITE-seq BMMC S4}. The CITE-seq BMMC S4 dataset contains simultaneous profiling of RNA expression and protein abundance from human BMMC cells sequenced by the CITE-seq protocol \cite{stoeckius2017simultaneous}. These cells came from 12 healthy human donors consisting of 3 batches in this dataset. The dataset encompasses 15,499 cells with measurements from 13,953 genes and 134 surface proteins.

\noindent\textbf{Brain 3k}. The human brain 3k dataset contains profiling of RNA expression from human healthy brain tissue sequenced by the 10x Genomics Multiome. The processed data include 3,233 cells with read counts for 36,601 genes.

\noindent\textbf{10x Multiome BMMC}. The 10x Multiome BMMC dataset contains simultaneous RNA expression and chromatin accessibility from human BMMC cells sequenced by the 10x Genomics Multiome protocol. Chromatin accessibility data were processed by dividing the genome into fixed-length bins, resulting in a 100,000-dimensional matrix. After filtering for gene expression, the gene expression matrix includes measurements for 20,000 genes per cell.

\section{Data availability}
All datasets used in this paper are publicly available. The SHARE-seq dataset is accessible in the GEO database under accession ID \href{https://www.ncbi.nlm.nih.gov/geo/query/acc.cgi?acc=GSE207308}{GSE207308}. The human brain dataset is available in the GEO database under accession ID \href{https://www.ncbi.nlm.nih.gov/geo/query/acc.cgi?acc=GSE214637}{GSE214637}. The 10x Multiome PBMC dataset is obtained from \href{https://scglue.readthedocs.io/en/latest/data.html}{https://scglue.readthedocs.io/en/latest/data.html}. The human fetal atlas dataset is obtained from \href{https://github.com/jsxlei/scCLIP}{https://github.com/jsxlei/scCLIP}. The CITE-seq BBMC S1 dataset, CITE-seq BBMC S4 dataset, and 10x Multiome BMMC dataset are available in the GEO database under accession ID \href{https://www.ncbi.nlm.nih.gov/geo/query/acc.cgi?acc=GSE194122}{GSE194122}. The human brain 3k dataset is available from \href{https://support.10xgenomics.com/single-cell-multiome-atac-gex/datasets/1.0.0/human_brain_3k}{https://support.10xgenomics.com/single-cell-multiome-atac-gex/datasets/1.0.0/human\_brain\_3k}.

\backmatter

\bibliography{sn-bibliography}

\end{document}